\documentclass{ws-procs975x65}

\begin{document}

\title{Linearised stability analysis of generic thin shells}

\author{Francisco S.~N.~Lobo$^1$, Prado Martin-Moruno$^2$, Nadiezhda Montelongo Garcia$^{3}$, Matt Visser$^2$}
\address{$^1$
Centro de Astronomia
e Astrof\'{\i}sica da Universidade de Lisboa, Campo Grande, \\
Edif\'{i}cio C8 1749-016 Lisboa, Portugal}
\address{$^2$ 
School of Mathematics, Statistics, and Operations Research,\\
Victoria University of Wellington, PO Box 600, Wellington 6140, New Zealand}
\address{$^3$
Departamento de F\'{\i}sica, Centro de Investigaci\'{o}n  y Estudios avanzados del I.P.N.,\\ 
A.P. 14-700,07000 M\'exico, DF, M\'exico}
%

\begin{abstract}

We construct generic spherically symmetric thin shells by using the 
cut-and-paste procedure. We take considerable effort to make the analysis as general and unified as practicable; investigating both the internal physics of the transition layer and its interaction with ``external forces'' arising due to interactions between the transition layer and the bulk spacetime. We demonstrate in full generality that stability of the thin shell is equivalent to choosing suitable properties for the material residing on the junction interface. Applications to gravastars and wormhole geometries are also explored.

\end{abstract}


\bodymatter

\section{Introduction}

We introduce a novel approach in the context of the linearised stability analysis of generic dynamical spherically symmetric thin shells \cite{MartinMoruno:2011rm,Garcia:2011aa}. The key point is to develop an extremely general and robust framework that can quickly be adapted to wide classes of generic thin-shell configurations. We consider standard General Relativity, with the transition layer confined to a thin shell. The bulk spacetimes (interior and exterior) on either side of the transition layer will be spherically symmetric and static but otherwise arbitrary. The thin shell (transition layer) will be permitted to move freely in the bulk spacetimes, permitting a fully dynamic analysis. This will then allow us to perform a general stability analysis, where the stability is related to the properties of the matter residing in the thin-shell transition layer.

\section{General formalism and linearised stability analysis}

To set the stage, consider two distinct spacetime manifolds, an \emph{exterior} ${\cal M_+}$, and an \emph{interior} ${\cal M_-}$, that are to be joined together across a surface layer $\Sigma$. 
Consider two generic static spherically symmetric spacetimes given by 
the following line elements:
\begin{eqnarray}
\hspace{-1.0cm}ds^2 = - e^{2\Phi_{\pm}(r_{\pm})}\left[1-\frac{b_{\pm}(r_{\pm})}
{r_{\pm}}\right] dt_{\pm}^2 + 
\left[1-\frac{b_{\pm}(r_{\pm})}{r_{\pm}}\right]^{-1}\,dr_{\pm}^2 + r_{\pm}^2 
d\Omega_{\pm}^{2}\,,
\label{generalmetric}
\end{eqnarray}
where $\pm$ refers to the exterior and interior geometry, respectively.

A single manifold ${\cal M}$ is obtained by gluing together ${\cal M_+}$ and ${\cal M_-}$ at their boundaries. Using the Darmois-Israel formalism, the surface stress components on the junction interface are given by
\begin{eqnarray}
\sigma&=&-\frac{1}{4\pi a}\left[
 \sqrt{1-\frac{b_{+}(a)}{a}+\dot{a}^{2}}
-\sqrt{1-\frac{b_{-}(a)}{a}+\dot{a}^{2}}
\right],
\label{gen-surfenergy2}
\\
{\cal P}&=&\frac{1}{8\pi a}\left[
\frac{1+\dot{a}^2+a\ddot{a}-\frac{b_+(a)+ab'_+(a)}{2a}}{\sqrt{1-\frac{b_{+}(a)}
{a}+\dot{a}^{2}}}     
+
\sqrt{1-\frac{b_{+}(a)}{a}+\dot{a}^{2}} \; a\Phi'_{+}(a)
\right. \nonumber\\
&&
\left. -
\frac{1+\dot{a}^2+a\ddot{a}-\frac{b_-(a)+ab'_-(a)}{2a}}{\sqrt{1-\frac{b_{-}(a)}
{a}+\dot{a}^{2}}}
-
\sqrt{1-\frac{b_{-}(a)}{a}+\dot{a}^{2}} \; a\Phi'_{-}(a)
\right].\label{gen-surfpressure2}
\end{eqnarray}
The surface mass of the thin shell is given by $m_s=4\pi a^2\sigma$.

The conservation equation is
	$\sigma'=-\frac{2}{a}\,(\sigma +{\cal P})+\Xi$,    
where $ \sigma'=d\sigma /da$. $\Xi$ is related to the energy-momentum flux that impinges on the shell, and is given by
\begin{equation}
\Xi =\frac{1}{4\pi a}\, \left[
\Phi_+'(a)\sqrt{1-\frac{b_+(a)}{a}+\dot{a}^{2}}
- 
\Phi_-'(a)\sqrt{1-\frac{b_-(a)}{a}+\dot{a}^{2}}
\right]\,.
\end{equation}
The flux term $\Xi$ is zero whenever $\Phi_\pm=0$; this is actually a quite common occurrence, for instance in either Schwarzschild or Reissner--Nordstr\"om geometries, or whenever $\rho+p_r=0$, so it is very easy for one to be mislead by those special cases.	


To analyse the stability of the thin shell, it is useful to rearrange the expression of $\sigma(a)$ into the form
$\frac{1}{2} \dot{a}^2+V(a)=0$,
where the potential $V(a)$ is given by
\begin{equation}
V(a)= {1\over2}\left\{ 1-\frac{b_{+}(a)+b_{-}(a)}{2a} -\left[\frac{m_{s}(a)}{2a}\right]^2-\left[\frac{b_{+}(a)-b_{-}(a)}{2m_{s}(a)}\right]^2\right\}\,.
   \label{potential}
\end{equation}

Note that $V(a)$ is a function of the surface mass $m_s(a)$. However, it is sometimes useful to reverse the logic flow and determine the surface mass as a function of the potential. Thus, if we specify $V(a)$, this tells us how much surface mass we need to put on the transition layer, which is implicitly making demands on the equation of state of the matter residing on the transition layer.

To analyse the stability of the thin shell, consider a linearisation around an assumed static solution, $a_0$, where one  Taylor expands $V(a)$ around $a_0$ to second order. Now, expanding around a static solution $\dot a_0=\ddot a_0 = 0$, we have 
$V(a_0)=V'(a_0)=0$, so it is sufficient to consider
$V(a)= \frac{1}{2}V''(a_0)(a-a_0)^2+O[(a-a_0)^3]$.   
The assumed static solution at $a_0$ is stable if and only if $V(a)$ has a 
local minimum at $a_0$, which requires $V''(a_{0})>0$. 

The condition $V''(a_{0})>0$ will be our primary criterion for the thin shell stability, though it will be useful to rephrase it in terms of more basic quantities. For instance, it is useful to express $m_s'(a)$ and $m_s''(a)$, which allows us to easily study linearised stability, and to develop a simple inequality on $m_s''(a_0)$ using the constraint $V''(a_0)>0$. Similar formulae hold for $\sigma'(a)$, $\sigma''(a)$, for ${\cal P}'(a)$, ${\cal P}''(a)$, and for $\Xi'(a)$, $\Xi''(a)$. In view of the redundancies coming from the relations $m_s(a) = 4\pi\sigma(a) a^2$ and the differential conservation law, the only interesting quantities are  $\Xi'(a)$, $\Xi''(a)$.

Thus, the stability condition $V''(a_0)\geq0$ is given as an inequality on $m_s''(a_0)$:
\begin{eqnarray}
\hspace{-1cm} m_s''(a_0) &\geq&
+{1\over4 a_0^3} 
\left\{ 
{ [b_+(a_0)- a_0 b_+'(a_0)]^2\over[1-b_+(a_0)/a_0]^{3/2}} 
- 
{ [b_-(a_0)- a_0 b_-'(a_0)]^2\over[1-b_-(a_0)/a_0]^{3/2}}
\right\}
\nonumber\\
&& 
+{1\over2} 
\left\{ 
{b_+''(a_0)\over\sqrt{1-b_+(a_0)/a_0}} 
-
{b_-''(a_0)\over\sqrt{1-b_-(a_0)/a_0}} \right\},
  \label{stable_ddms1}
\end{eqnarray}
provided $b_+(a_0)\geq b_-(a_0)$. If $b_+(a_0)\leq b_-(a_0)$ the direction of the inequality is reversed. In the absence of external forces this inequality is the only stability constraint one requires.
However, once one has external forces ($\Xi\neq 0$), we have a second stability inequality:
\begin{eqnarray}
\hspace{-1.25cm}
\left.[4\pi\,\Xi(a)\,a]''\right|_{a_0} &\leq& \left.\left\{ 
\Phi_+'''(a) \sqrt{1-b_+(a)/a} - 
\Phi_-'''(a) \sqrt{1-b_-(a)/a} \right\}\right|_{a_0}
\nonumber\\
&& 
\hspace{-2.0cm}
- \left.\left\{ 
\Phi_+''(a) { (b_+(a)/a)'\over\sqrt{1-b_+(a)/a}} - 
\Phi_-''(a){(b_-(a)/a)'\over\sqrt{1-b_-(a)/a}} \right\}\right|_{a_0}
\nonumber\\
&&
\hspace{-2.0cm}
-{1\over4} 
\left.\left\{ 
\Phi_+'(a) { [(b_+(a)/a)']^2\over[1-b_+(a)/a]^{3/2}} -
\Phi_-'(a) {[(b_-(a)/a)']^2\over[1-b_-(a)/a]^{3/2}} \right\}\right|_{a_0}
\nonumber\\
&&
\hspace{-2.0cm}
-{1\over2} 
\left.\left\{ 
\Phi_+'(a) { (b_+(a)/a)''\over\sqrt{1-b_+(a)/a}} -
\Phi_-'(a) {(b_-(a)/a)''\over\sqrt{1-b_-(a)/a}} \right\}\right|_{a_0},
   \label{stability_Xi}
\end{eqnarray}
if $\Phi'_+(a_0)/\sqrt{1-b_+(a_0)/a_0} \geq \Phi'_-(a_0)/\sqrt{1-b_-(a_0)/a_0} $. Note that this last equation is entirely vacuous in the absence of external forces.

\section{Conclusion}

We have developed a novel and extremely general and robust framework leading to the linearised stability analysis of dynamical spherically symmetric thin shells. A key point in the conservation law of the surface stresses is the presence of a flux term. More specifically, we have considered the surface mass as a function of the potential, so that specifying the latter tells us how much surface mass we need to put on the transition layer. This procedure demonstrates in full generality that the stability of the thin shell is equivalent to choosing suitable properties for the material residing on the thin shell.

\section*{Acknowledgements}
FSNL acknowledges the CERN/FP/123615/2011 and CERN/FP/123618/2011 FCT grants.
PMM thanks financial support from the Spanish Ministry of Education through a FECYT grant, via the postdoctoral mobility contract EX2010-0854.


\begin{thebibliography}{99}

\bibitem{MartinMoruno:2011rm} 
  P.~Martin Moruno, N.~Montelongo Garcia, F.~S.~N.~Lobo and M.~Visser,
  JCAP {\bf 1203}, 034 (2012).  

\bibitem{Garcia:2011aa} 
  N.~M.~Garcia, F.~S.~N.~Lobo and M.~Visser,
  Phys.\ Rev.\ D {\bf 86}, 044026 (2012).  


\end{thebibliography}
\end{document}